\documentclass{elsart}

\usepackage{graphicx}
\usepackage{natbib}
\usepackage{amssymb}

\begin{document}

\bibliographystyle{unsrt}

\begin{frontmatter}

\title{Stability diagrams for bursting neurons modeled by
three-variable maps}

\author[pernambuco]{M. Copelli\corauthref{cor}},
\corauth[cor]{Corresponding author}
\ead{mcopelli@df.ufpe.br}
\author[floripa]{M. H. R. Tragtenberg},
\author[ribpret]{O. Kinouchi}

\address[pernambuco]{
Laborat\'orio de F\'{\i}sica Te\'orica e Computacional \\
Departamento de F\'{\i}sica, Universidade Federal de Pernambuco \\
50670-901 Recife, PE, Brazil
}
\address[floripa]{
Departamento de F\'{\i}sica, Universidade Federal de Santa
Catarina \\
88040-900 Florian\'opolis, SC, Brazil
}
\address[ribpret]{
Departamento de F\'{\i}sica e Matem\'atica,\\
Faculdade de Filosofia, Ci\^encias e Letras de Ribeir\~ao Preto, \\
Universidade de S\~ao Paulo\\
Av. dos Bandeirantes 3900, 14040-901 Ribeir\~ao Preto, SP, Brazil
}

\begin{abstract}
We study a simple map as a minimal model of excitable cells. The map
has two fast variables which mimic the behavior of class I neurons,
undergoing a sub-critical Hopf bifurcation. Adding a third slow
variable allows the system to present bursts and other interesting
biological behaviors. Bifurcation lines which locate the excitability
region are obtained for different planes in parameter space.
\end{abstract}


\end{frontmatter}

\section{Introduction}
\label{intro}

Maps are dynamical systems with continuous state variables but
discrete-time dynamics. They constitute low computational cost
elements for large scale simulations of complex
systems~\cite{Aihara90,Kaneko94,Kaneko00}. The proper use of coupled
map lattices (CML) in neural modeling depends on the choice of a good
basic element. Two-variable maps have been proposed by Aihara
et. al.~\cite{Aihara90}, Chialvo~\cite{Chialvo95}, and Kinouchi and
Tragtenberg~\cite{Kinouchi96b}. Methodological considerations about
the use of maps in computational neuroscience can be found in these
references.

A three-variable map with one slow and two fast variables has been
recently introduced by Kuva et al.~\cite{Kuva01}. It is able to
reproduce complex biological behaviors, such as bursting, cardiac-like
spikes, chaotic orbits and slow regular spikes (``singlet bursting'').
Here we determine stability diagrams for this map, which may be
thought of as a discrete-time analogous of the Hindmarsh-Rose
neuron~\cite{Hindmarsh84}. We focus exclusively on the lines at which
the fixed point loses stability, which define the limit of the
excitability region. We extend previous results by obtaining diagrams
beyond the adiabatic limit, where the slow variable has a much slower
time scale than the other two~\cite{Kuva01}.

\section{The model}
\label{themodel}

The following three-dimensional non-linear map has been proposed as a
minimal excitable bursting cell model~\cite{Kuva01}:
\begin{eqnarray}
x(t+1) &=& \tanh\left(\frac{x(t) - K y(t) + z(t) + I(t) }{T} 
\right) \:, \nonumber \\
y(t+1) & = & x(t) \label{yosmap} \:,\\
z(t+1) & = & \left(1-\delta \right) z(t) - \lambda
\left(x(t)-x_R\right)\; .
\label{zevol} \nonumber
\end{eqnarray}
Like in the Hindmarsh-Rose model~\cite{Hindmarsh84}, the variable $x$
represents the instantaneous membrane potential, $y$ is a recovery
variable and $z$ can be considered a slow adaptive current if $\delta$
and $\lambda$ are set to small values ($\ll 1$). Under these
conditions, we can identify $z(t)$ with e.g.  the Ca$^{2+}$-dependent
K$^+$ current $I_{AHP}$~\cite{Rinzel98b,Koch}, since its time scale
and general effect (long lasting hyperpolarizaton and adaptation of
firing rates) are very similar. $I(t)$ represents an external current
which can model the effects of a synapse, a stimulus-induced ionic
flow in sensory cells, or an experimentally controlled clamped
current.

\begin{figure}[bt]
\begin{center}
\includegraphics[width=0.4\columnwidth]{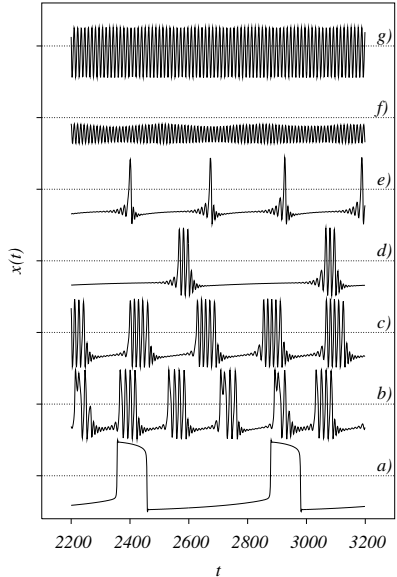} 
\caption{\label{fig:waves}Different dynamical behaviors of the
three-dimensional map for $K=0.6$. Unless otherwise stated, $T=0.35$
and $\delta=\lambda=0.001$: a) cardiac-like spikes (spikes with
plateau, $T=0.25$, $x_R=-0.5$); b) chaotic behavior ($T=0.322$,
$x_R=-0.4$); bursting behavior in c) ($x_R=-0.45$) and d)
($x_R=-0.6$); e) regular spiking or ``singlet bursting''
($\delta=\lambda=0.003$, $x_R=-0.62$); f) sub-threshold oscillations
($T=0.45$, $x_R=-0.5$) and g) fast spiking ($T=0.45$, $x_R=-0.2$). For
better visualization, the curves have been shifted vertically by 2
(dimensionless) units. The dotted lines correspond to $x=0$.}
\end{center}
\end{figure}

If we set $z(t)=H=\mbox{const}$, we are left with the fast dynamical
system $(x,y)$ which has been studied in Ref.~\cite{Kinouchi96b}. The
analysis of the two-variable system shows~\cite{Kinouchi96b} that,
with $T$ and $K$ fixed, a sub-critical Hopf bifurcation occurs if we
increase the term $H+I$ (that is, the bias parameter $H$ and the
external current $I$). With $I=0$, the autonomous behavior is
controlled by the bias term $H$ and changes dramatically at the
bifurcation value $H_c^\pm=T\mbox{atanh}(x_c^\pm) - (1-K)x_c^\pm$,
where $x_c^\pm=\pm\sqrt{1-T/K}$.  

Close to the bifurcation line, the behavior is that of an excitable
cell~\cite{Kinouchi96b}.  Since the Hopf bifurcation is sub-critical
(``hard excitation''), this model belongs to the class of the
FitzHugh-Nagumo and Hodgkin-Huxley models (class I
neurons)~\cite{FitzHugh60,Koch}, where the stable fixed point is a
focus (perturbations lead to spiraling behavior in the phase
plane). These small oscillations are described in biophysical models,
after linearization about the fixed point, as a phenomenological
inductance behavior (see chapter 10 in~\cite{Koch}).

\section{Stability Diagrams for the Three-Dimensional Map}
\label{results}

The three-variable model presents several qualitatively different
dynamical behaviors which are usual in real biological neurons
(Fig.~\ref{fig:waves}). In particular, the slow current $z(t)$ is
responsible for the bursting dynamics, as discussed by Rinzel and
Ermentrout~\cite{Rinzel98b}.  This slow dynamics is controlled by the
parameters $(\delta,\lambda,x_R)$. In what follows, we focus on the
analysis of the stability diagrams in terms of these parameters. The
lines where the fixed point $(x_*,y_*,z_*)$ loses stability (for
$I=0$) have been determined by standard linear stability analysis,
which yields the equation
\begin{equation}
-\Lambda^3+[\alpha+(1-\delta)]\Lambda^2-\alpha[\lambda+K+1-\delta]\Lambda
+ K\alpha(1-\delta) = 0
\end{equation}
for the eigenvalues $\Lambda$, where
$\alpha=(1-x_*^2)/T$.

\subsection{The case $\delta=0$ \label{delta=0}}

\begin{figure}[!tb]
\begin{center}
\includegraphics[width=0.47\columnwidth]{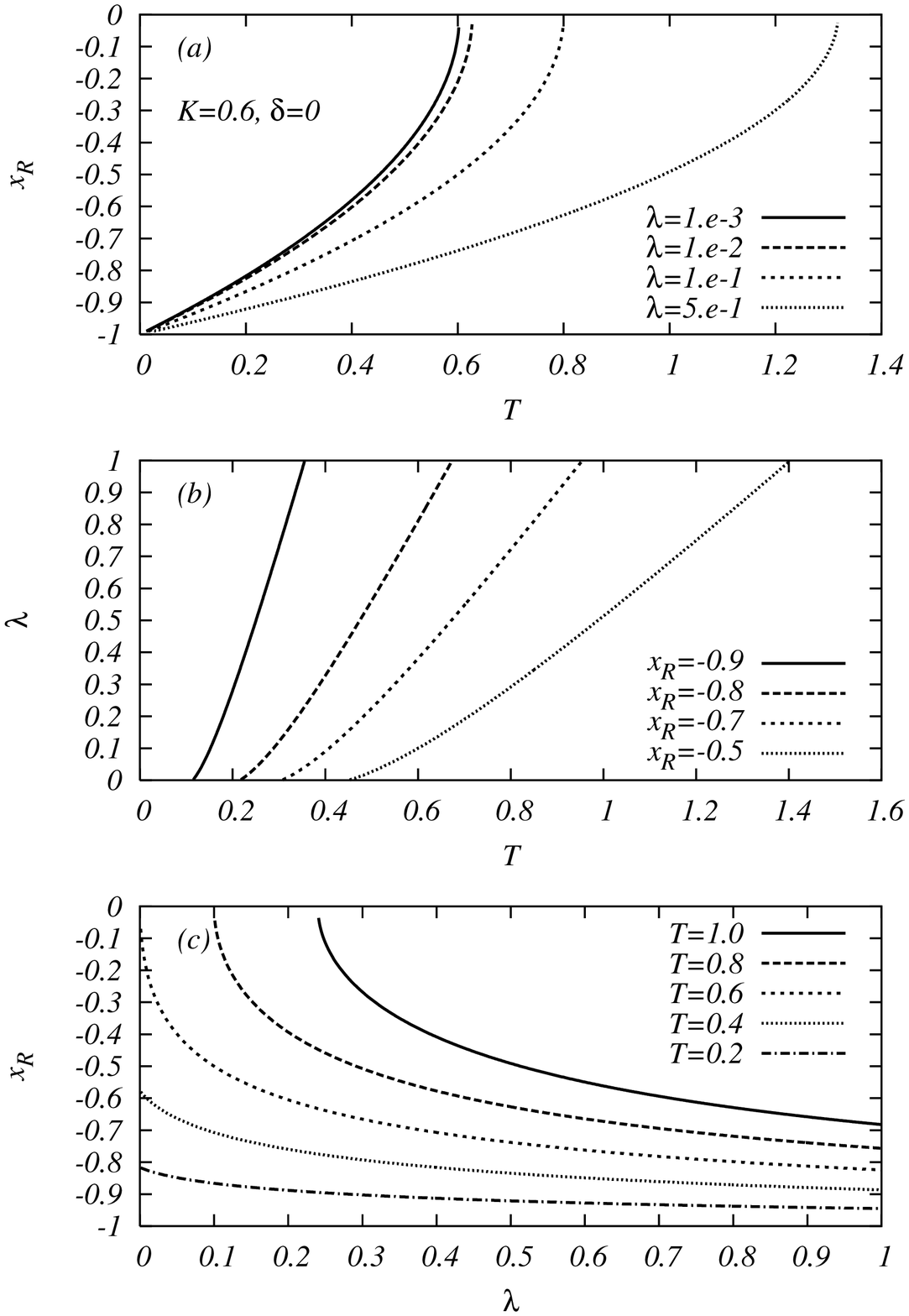} 
\includegraphics[width=0.47\columnwidth]{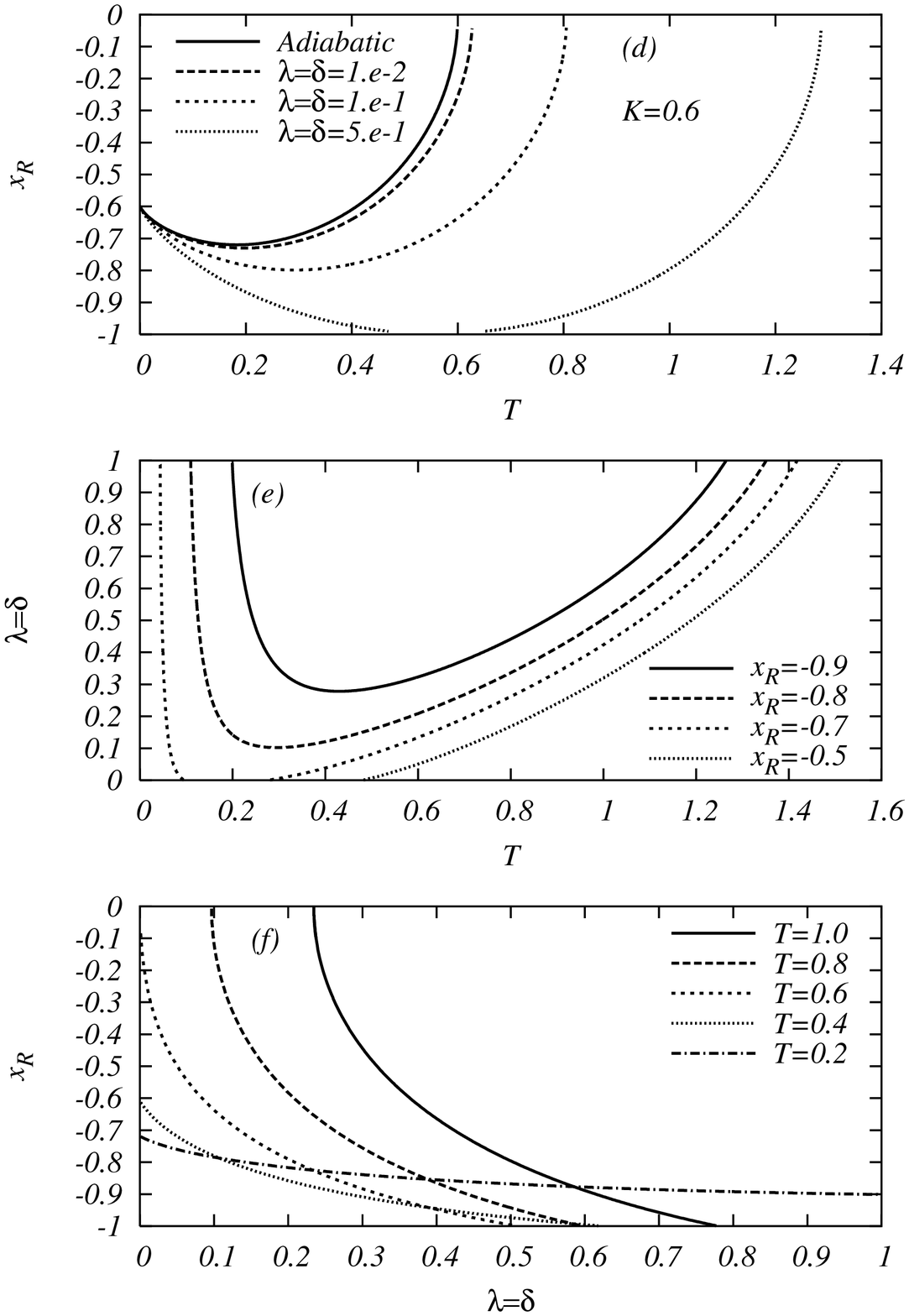}
\caption{\label{fig:delta=0} Left panel: bifurcations for
$\delta=0$. Right panel: bifurcations for $\lambda=\delta$. The fixed
point is stable below the lines. }
\end{center}
\end{figure}

The case $\delta=0$ provides a limiting behavior where the neuron can
be ``perfectly adaptive''. This means that, when the fixed point is
stable, adding an external current $I$ produces a transient response:
firing only occurs while $z(t)$ is adapting (phasic
firing)~\cite{Kuva01}.  The fixed points can be determined exactly in
that case: $x_*=y_*=x_R$ and $z_*=T\mbox{atanh}(x_R)-(1-K)x_R$.  This
system presents monotonic behavior as far as loss of stability is
concerned, as shown in the left panel of Fig.~\ref{fig:delta=0}: with
all the other parameters frozen, the fixed point loses stability by
either increasing $x_R$, increasing $\lambda$ or decreasing $T$.

\subsection{The case $\lambda=\delta$ \label{lambda=delta}}

The subspace $\lambda=\delta$ reflects the simple case where the time
scales of the inflow ($\lambda$) and outflow ($\delta$) of the $z$
current are the same. As opposed to the $\delta=0$ case, the
bifurcation line in the $x_R$ vs. $T$ plane now presents a minimum,
signaling that for sufficiently low values of $x_R$, the fixed point
can become stable by lowering $T$ (Fig.~\ref{fig:delta=0}d). This
non-monotonic behavior in the parameter $T$ is also observed in the
planes $(T,\lambda)$ and $(\lambda,x_R)$ (Figs.~\ref{fig:delta=0}e
and~\ref{fig:delta=0}f).

\subsection{Adiabatic approximation}
\label{adiabatic}

In the limit of $\delta,\lambda \ll 1$ one can obtain analytical
results by means of an adiabatic approximation. By taking $z$ to be a
quasi-static variable, in the fixed point region the stationary $z_*$
value is determined by $z_* = (\lambda/\delta) (x_R-x_*)$. It is thus
clear that the parameter $x_R$ is essential for controlling the
equilibrium value of $z$. Applying the results for the fast
sub-system~\cite{Kinouchi96b}, the adiabatic approximation
$(z_*)_c=H_c$ defines a critical value for the reversion potential:
$x_R^c(T) = x_c^\pm+ (\delta /\lambda)H_c^\pm = \pm
\left[1-(\delta/\lambda)(1-K)\right] \sqrt{1-T/K} \pm (\delta/\lambda)
T \:\mbox{atanh}\left(\sqrt{1-T/K} \right)$. This critical line is
plotted in the $(x_R,T)$ plane for $K=0.6$ (Fig.~\ref{fig:delta=0}d),
and is visually indistinguishable from the results for
$\lambda=\delta=0.001$.

\subsection{The ($I$,$T$) plane}

As observed in the previous section, within the adiabatic
approximation the critical current $I_c$ for the three-variable model
can be readily obtained from the two-dimensional result, $I_c= H_c
-z_* = H_c(T) +(\lambda/\delta)(x_c(T)-x_R) $, with
$H_c=\mbox{min}\{H_c^+,H_c^-\}$ and $x_c= \min\{x_c^+,x_c^-\}$.
Remembering that the critical $x_R^c$ for $I=0$ is $x_c +
(\delta/\lambda) H_c $, we finally get $I_c(T) =
(\lambda/\delta)(x_R^c(T)-x_R)$.  This means that the fixed point
stability boundary $I_c(T)$ is exactly the same as the $x_R$ versus
$T$ diagram (Fig.~\ref{fig:delta=0}d) with a linear shift in the $I$
scale given by the above equation.  For example, if we set
$\delta=\lambda\ll1$  and $x_R=-1$, the element starts firing at $I_c(T)=
x_R^c(T) + 1$.

\begin{figure}[!bt]
\begin{center}
\includegraphics[width=0.6\columnwidth]{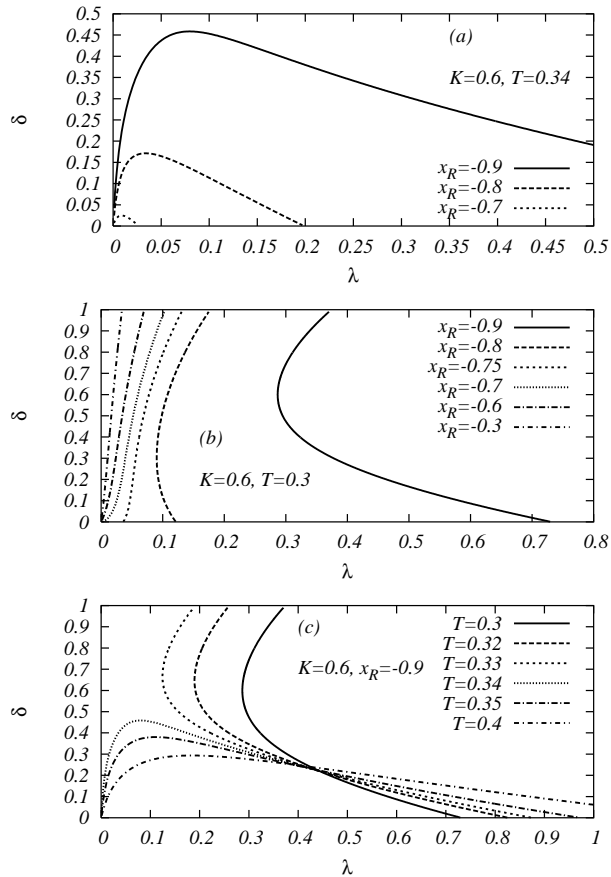}
\caption{\label{fig:lambdavsdelta}Hopf bifurcations in the
$(\lambda,\delta)$ plane for $K=0.6$. The fixed point is stable (a)
under the lines; (b) to the left of the lines. (c) Evolution of the
curves from (a) to (b).}
\end{center}
\end{figure}

\subsection{The $(\lambda$,$\delta)$ plane \label{deltavslambda}}

Relaxing the constraint $\lambda=\delta$ and moving away from the
adiabatic condition $\lambda,\delta\ll 1$ gives rise to an interesting
interplay between $\delta$, $\lambda$ and
$T$. Fig.~\ref{fig:lambdavsdelta}a shows the $\delta$ vs $\lambda$
stability boundaries for $T=0.34$. While the effect of $x_R$ remains
monotonic (i.e. larger values of $x_R$ tend to destabilize the fixed
points), the curves $\delta_c(\lambda)$ have a maximum. An interesting
phenomenon occurs if $T$ is lowered (Fig.~\ref{fig:lambdavsdelta}b,
with $T=0.3$): the shape of the boundaries changes dramatically, with
larger values of $\lambda$ inducing the instability of the fixed
point. Fig.~\ref{fig:lambdavsdelta}c shows how the curves change as
$T$ varies. For larger values of $T$, the stability of the system is
weakly dependent on $\lambda$, depending rather on sufficiently small
values of $\delta$. The change in the shapes can be intuitively
understood on the basis of Eqs.~\ref{yosmap}: large $T$ means small
increments in the values of $x$, rendering the $z$-equation strongly
dependent on $\delta$ (Fig.~\ref{fig:lambdavsdelta}a). For small $T$,
on the other hand, $\delta$ becomes less relevant: since the changes
in $x$ are larger, $\lambda$ must be small for the fixed point to
remain stable (Fig.~\ref{fig:lambdavsdelta}b).

\section{Conclusions}
\label{conclusions}

We studied a three variable nonlinear map which includes a slow
dynamics over a sub-critical Hopf bifurcation generically found in
class I neurons. This element is used to represent different behaviors
observed in biological cells such as excitability, regular spiking,
spikes with plateau and bursts~\cite{Kuva01}. We have presented
bifurcation lines in different planes in the parameter space. This
gives us a global understanding of the dynamics of the map and enables
the proper choice of parameters for locating the excitability
region. These maps can be coupled by properly modeling chemical or
electrical synapses~\cite{Kuva01}. We are currently applying this
coupled map framework to large scale biologically realistic
architectures, results will be published elsewhere.

\section*{Acknowledgments}
It is a pleasure to acknowledge useful discussions with A. C. Roque,
J. P. Neirotti, N. Caticha, S. M. Kuva and R. F. Oliveira. We
acknowledge support from CNPq, FACEPE, Projeto Enxoval-UFPE and
FAPESP. We thank S. R. A. Salinas for the hospitality of the
Statistical Mechanics group at IFUSP, where part of this manuscript
was written.


\begin{thebibliography}{10}

\bibitem{Aihara90}
K.~Aihara, T.~Takabe, and M.~Toyoda.
\newblock Chaotic neural networks.
\newblock {\em Phys. Lett. A}, 144:333--340, 1990.

\bibitem{Kaneko94}
K.~Kaneko.
\newblock Relevance of dynamic clustering to biological networks.
\newblock {\em Physica D}, 75:55--73, 1994.

\bibitem{Kaneko00}
K.~Kaneko and I.~Tsuda.
\newblock {\em Complex Systems: Chaos and Beyond, A Constructive Approach with
  Applications in Life Sciences}.
\newblock Springer Verlag, 2000.

\bibitem{Chialvo95}
D.~R. Chialvo.
\newblock Generic excitable dynamics on a two-dimensional map.
\newblock {\em Chaos Soliton Fract.}, 5:461--480, 1995.

\bibitem{Kinouchi96b}
O.~Kinouchi and M.~H.~R. Tragtenberg.
\newblock Modeling neurons by simple maps.
\newblock {\em Int. J. Bifurcat. Chaos}, 6(12A):2343--3460, 1996.

\bibitem{Kuva01}
S.~M. Kuva, G.~F. Lima, O.~Kinouchi, M.~H.~R. Tragtenberg, and A.~C. Roque.
\newblock A minimal model for excitable and bursting elements.
\newblock {\em Neurocomputing}, 38-40:255--261, 2001.

\bibitem{Hindmarsh84}
J.~L. Hindmarsh and R.~M. Rose.
\newblock A model of neuronal bursting using three coupled first order
  differential equations.
\newblock {\em Proc. Roy. Soc. London B}, 221:87--102, 1984.

\bibitem{Rinzel98b}
J.~Rinzel and B.~Ermentrout.
\newblock Analysis of neural excitability and oscillations.
\newblock In C.~Koch and I.~Segev, editors, {\em Methods in Neuronal Modeling:
  From Ions to Networks}, pages 251--292. MIT Press, 2nd edition, 1998.

\bibitem{Koch}
Christof Koch.
\newblock {\em Biophysics of Computation}.
\newblock Oxford University Press, New York, 1999.

\bibitem{FitzHugh60}
R.~FitzHugh.
\newblock Thresholds and plateaus in the {H}odgkin-{H}uxley nerve equations.
\newblock {\em J. Gen. Physiol.}, 43:867--896, 1960.

\end{thebibliography}

\end{document}